\documentclass{article}

\usepackage{cite}

  \usepackage[pdftex]{graphicx}
  \graphicspath{{../img/}}

\usepackage{algorithmic}

\usepackage{array}

\usepackage{url}

\begin{document}

\title{{\sf NodIO}, a JavaScript framework for volunteer-based evolutionary algorithms : first results}

%(Mario) Maybe:  {\sf NodIO}, a JavaScript framework for
%volunteer-based evolutionary algorithms.
% Accepted with a small change :-) - JJ

\author{Juan-J.~Merelo*$^1$, Mario Garc\'ia-Valdez$^2$, Pedro
  A. Castillo$^1$,\\ 
Pablo Garc\'ia-S\'anchez$^1$, P. de las Cuevas*$^1$, Nuria Rico$^3$
\thanks{Manuscript submitted for review on \today.}%
\thanks{$^1$Dept. of Computer Architecture and Technology and CITIC University of Granada}%
\thanks{$^2$Dept. of Graduate Studies at Instituto Tecnol\'ogico de Tijuana}%
\thanks{$^3$Dept. of Statistics and Operational Research University of Granada}%
\thanks{E-mail addresses: {\tt jmerelo@ugr.es}, {\tt
    mario@tectijuana.edu.mx}, {\tt pacv@ugr.es}, {\tt pablogarcia@ugr.es},{\tt nrico@ugr.es}}%
\thanks{*Corresponding author.}%
}

\maketitle

\begin{abstract}

JavaScript is an interpreted language mainly known for its inclusion
in web browsers, making them a container for rich Internet based applications. This has
inspired its use, for a long time, as a tool for evolutionary
algorithms, mainly so in browser-based volunteer computing
environments. Several libraries have also been published so far and are in
use. However, the last years have seen a resurgence of interest in the
language, becoming one of the most popular and thus spawning the
improvement of its implementations, which are now the foundation of many
new client-server applications. We present such an application for
running distributed volunteer-based evolutionary algorithm 
experiments, and we make a series of
measurements to establish the speed of JavaScript in evolutionary
algorithms that can serve as a baseline for comparison with other
distributed computing experiments. These experiments use different
integer and floating point problems, and prove that the speed of
JavaScript is actually competitive with other languages commonly used
by the evolutionary algorithm practitioner. 

\end{abstract}


%---------------------------------------------------------------
\section{Introduction}

The need to use non-permanent and potentially massive computing
infrastructure for distributed scientific experiments has lead to the
rising of {\em volunteer computing} defined as a set of tools that allow
citizens to {\em donate} cycles to scientific experiments by running
an application \cite{Anderson06thecomputational}; this application can be embedded in a web page. And
since browsers are ubiquitous, we are more interested in these kind of systems, this has led us to create a framework for
evolutionary algorithm experiments that is able to work in that mode
\cite{DBLP:journals/corr/GuervosG15}. {\sf NodIO} is a
cloud or bare metal based volunteer evolutionary computing system
derived from the {\sf NodEO} library, whose architecture has been
developed using JavaScript on the client as well as the server.
All parts of the framework are free and available with a free license
from \url{https://github.com/JJ/splash-volunteer}.

%%% Issues with volunteer based systems

As a volunteer based system, there are some issues that have to be addressed 
\cite{sarmenta2001volunteer,web:BOINC} :
\begin{enumerate}
\item Volunteers are anonymous, only the Internet Protocol addresses (IP) sent in 
each request is known.
\item As anonymous entities, volunteers are not accountable. 
If a volunteer misbehaves in some way the provider cannot 
prosecute or discipline the volunteer. In the case of JavaScript,
this issue is aggravated because even if it is obfuscated, source code
and data is sent to clients. 
\item Participants must also trust the application provider, 
regarding invasion of privacy, the integrity of their devices, 
and intended application. 
\end{enumerate}

Several kind of threats are possible in this environment; first, since its
programming interface is open and its code is open source, it is
relatively easy to find vulnerabilities and 
sabotage the system, as indicated in \cite{domingues2007sabotage}, by
crafting a fake request which, for instance, assigns a fake fitness to
a particular chromosome.
Other two different kinds of attacks are also
possible: a denial of service attack as well as penetration
testing. We have approached this threat in a pragmatic way: all
sources, client and server, for the application, are openly released
as free software. 
 We also released as an open access paper the result of the 
 experiments performed \cite{DBLP:journals/corr/GuervosG15}  
 and all data, once anonymized, as open data. This
builds trust between the user and the scientist, which indeed has an
impact on the performance of the system, since there is no need to
include cheating checks or other functions that would degrade it. This
socio-technical approach is, indeed, 
coherent with our whole approach to the study of a distributed volunteer
 system which can only be tackled by approaching it from the
technical as well as the social, as in social media, angle. 

% Justification for JavaScript
In general, a distributed volunteer-based evolutionary computation
framework based on the browser such as NodIO is simply a client-server system
whose client is, or can be, embedded in the browser via
JavaScript. Since JavaScript is  the only language that is present
across all browsers, the choice was quite clear. Even so, there might
be some issues with the performance of the language itself, which 
 is
why we have made a comparison between JavaScript and other languages
\cite{2015arXiv151101088M}. In that paper we prove that JavaScript can be
faster than compiled languages such as Scala and, in any case, it provides
a performance that is comparable to other languages usually employed % (Paloma) Adverb
in evolutionary computation and numerical optimization such as Python. 
Compiled languages such as
Java or C might be faster when 
considering the performance of a single-user, 
but this is more than offset by the computational power of
the spontaneous volunteers we can gather at whim, including people
using their mobile phones or tablets. Together, the performance is several orders of magnitude
higher, which is the objective in this kind of systems.

Since JavaScript is
quite a popular language nowadays, it also provides compilers from a number
of languages, many of them in their family (like CoffeeScript), but
many other outside, so researchers can, in fact, write their fitness
function in Python, Erlang, Scala, or even Java, and then compile it to
JavaScript. For completeness, an experiment comparing the performance of
a JavaScript implementation versus Java and Matlab is included in section
\ref{sec:experiments}. We will first describe the {\sf NodIO} architecture in
the next Section.

\section{{\sf NodIO } architecture}

\begin{figure}[!t]
\centering
\includegraphics[width=4in]{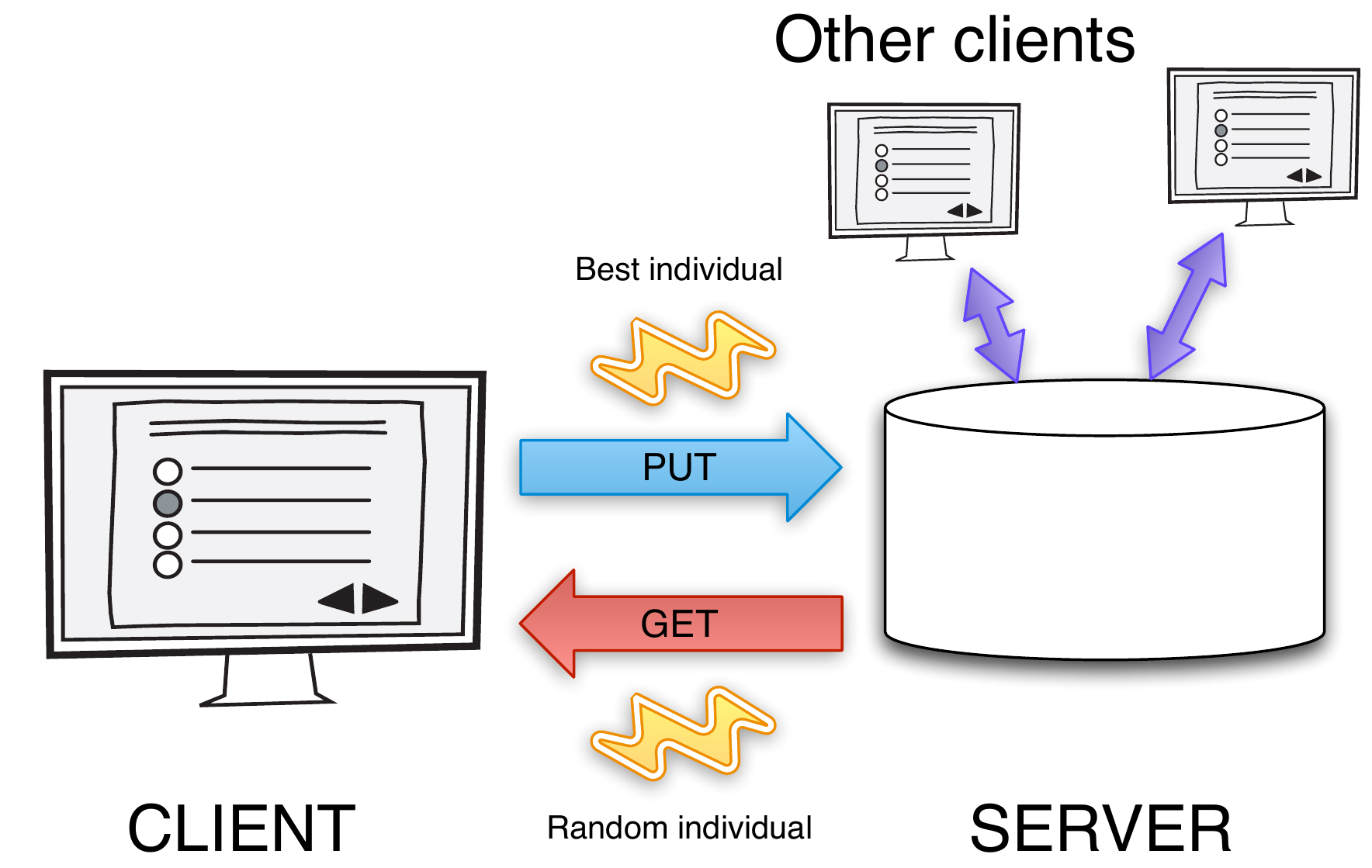}
\caption{Description of the proposed framework. Clients execute a JavaScript EA
  in the browser, which, every 100 generations, sends the best
  individual and receives a random one from the server.}
\label{fig:system}
\end{figure}

The {\sf NodIO} architecture has two tiers. 
The figure \ref{fig:system} describes the general system architecture and
algorithm behavior. Different web technologies, such as JQuery or {\tt
  Chart.js} have
been used to build the user interface elements of the framework.
\begin{enumerate}

\item A REST server, that is, a server that includes several {\em
  routes} 
  that can be called for storing and retrieving information (the ``CRUD'' cycle:
  create, request, update, and delete) from the server. 
  A JSON data format is used for the communication between 
  clients and the server. There are two kinds of information:
  {\em problem} based, that is, related to the
  evolutionary algorithm such as {\tt PUT}ing a chromosome in or {\tt
  GET}ing a random chromosome from it, and {\em information} related
  to the performance and state of the experiments. It also performs logging
  duties, but they are basically a very lightweight and high performance
  data storage \cite{jj:idc:lowcost}.
  The server has the capability to
  run a single experiment, storing the chromosomes in a data structure
  that is reset when the solution is found.

\item A client that includes the evolutionary algorithm as
  JavaScript code embedded in a web page that displays graphs, some
  additional links, and information on the experiment. This code runs
  an evolutionary algorithm {\em island} starting with a random
  population, then it sends, every 100 generations, the best individual
  back to the server (via a {\tt PUT} request), and requests a random
  individual from the server (via a {\tt GET} request). % (Paloma) Re-writing because there are 3 steps (island creation -> PUT -> GET)
\end{enumerate}

  This is the invariant part of the framework, but other than that,
  the algorithm can be run in many different ways: 
  stopping when a solution is found or using Web Workers. In fact, since it is a pool-based system such as
  the one described in \cite{LNCS86720702}, any kind of client that
  calls the application programming interface (API) can be used,
  written in any kind of language. However, in this paper we are going
  to focus on the dynamics and measurements of the volunteer part of
  the framework. 

The researcher only has to change a function, which can be written in
JavaScript or in any other language with cross-compilation to
JavaScript \cite{web:compilersjs} to solve different optimization
problems. 

Let us see how this system addresses the different challenges
outlined above:
\begin{itemize}
\item {\em Scalability} is provided via the use of a lightweight and
  high-performance, single-threaded, server based in Node.js and
  Express.js. Although this single server is a bottleneck since it
  will eventually saturate, the fact that it runs as a non-blocking single thread
  allows the service of many requests. In fact, a limit in the
  number of simultaneous requests will be reached, but so far it has
  not been found, unlike what we found in our previous systems, DCoR,
  \cite{gecco07:workshop:dcor}, which had a low scalability. 
\item The framework is {\em heterogeneous} since it does not need any
  performance, operating system, or even browser requirement as long
  as JavaScript is enabled: anyone
  visiting the page, even from mobile devices, can load the algorithm.
  In case the browser does not support HTML5 Web workers (e.g., 
  Windows Phone or some version of Android) or their use
  has been disabled, a basic version of {\sf NodIO} can also be used.
\item {\em Fault tolerance} is always an issue, and in this case, the
  single point of failure would be the server: the system, as such,
  would break down if the server fails. However, the individual
  islands in every browser would continue running, and having access
  to just one of them would allow the local algorithm to proceed. In
  fact, the island does not need the server to run: it runs locally if
  needed, with the only exception that it is obviously unable to
  communicate with the rest of the islands.
\item {\em Adaptiveness} is achieved simply through the autonomous
  operation of every individual island without any synchronization
  mechanism. The islands in the system are, in fact, unaware of each
  other, communicating only through the server. This architecture 
  has been tested in other high demand systems with success, but 
  in the case of {\sf NodIO}, additional experiments are needed to asses 
  the scalability of the communication system. % (Paloma) Is it ok if
                                % we talk about scalability when we
                                % were talking about adaptiveness?
                                % Scalability was also described in
                                % the first point. 
  %(Mario) Scalability of the communication system in particular?
\item Since the algorithm runs in the browser, user's {\em safety} is
  achieved through its sandbox mechanisms. The user is thus assured
  that there is no unsafe access either to their local files or even
  to more resources than the browser should be allotted.
\item Running the algorithm is just a matter of loading the page,
  which makes the operation totally {\em anonymous}. For the same
  reason, {\em ease of use} is optimal, being as easy as simply
  clicking on an URL, available to anyone with access to a browser.
\item {\em Reasonable performance} is not ensured. In fact, we should
  make sure that there is a reasonable amount of clients over which
  the performance achieved is better than what you would obtain in
  your own desktop system. If this is not the case, it is a pointless
  academic exercise.
\end{itemize}

There are many different ways to validate a framework that intends to
address these challenges. Some of them are included by design: it is
anonymous, since no accounts are needed to participate in the
experiment, it is safe for the user, since we are using the browser
black box, and it is also heterogeneous. We have designed the
experiments so that they show adaptability to different types of
experiments and clients, scalability with the number of clients and
the problem size, and measured performance by comparing it with
a baseline configuration, with which, as indicated above, is
compatible and can be used in a complementary way. Fault tolerance
will be experimented by dropping the server and checking the continuity
of the experiments. 

After the initial experiments \cite{DBLP:journals/corr/GuervosG15}, we
realized that we needed to make some improvements. So, in order to
improve the number of cycles per user, we enhanced the 
basic architecture in several different aspects. One proposed improvement
involved restarting the client once the solution was
found, so that as long as the user was visiting the page it was
running an evolutionary algorithm. Another was a small algorithmic
enhancement: instead of having a fixed population,
population size was randomly distributed between 128 and 256 and
different for each client. We called this instance of
the architecture NodIO-W$^2$. The other improvement was to 
use the Web Worker API as described next.

The rationale for this change comes from the realization that JavaScript has been commonly used to develop client-side scripts
that handle the user interface, communicating asynchronously with servers and
updating the content that is displayed \cite{flanagan2006javascript}.
Long-running scripts like those proposed in this work can interfere with the
responsiveness of the web application, because everything is handled by a
single thread. The Web Workers specification \cite{hickson2012web} defines an
API that allows Web developers to execute long-running scripts in parallel
with the scripts of the main page. Worker scripts are spawned in the
background allowing a thread-like operation using message-passing as the
communication mechanism between the worker's execution environment and the
main thread. According to the Web Workers W3C Candidate Recommendation
\cite{hickson2012web} workers are expected to be long-lived, they have a high
start-up performance cost, and a high per-instance memory cost. The Web Worker
API is implemented in current versions of both desktop and mobile web browsers.
For volunteer computing using the Web Worker API brings several advantages
over a single threaded implementation:

\begin{itemize}
\item If the browser uses a tabbed document interface, the worker script
keeps running in the background even in case the user brings forward another tabbed
window.
\item Several evolutionary algorithms can be executed in parallel in a single web
page. 
\item Many implementations of the Web Worker API can use multiple-core CPUs for
the execution of worker scripts, in particular the execution in multiple-core CPUs
was tested in desktop versions of Chrome, FireFox and Safari.
\item Each {\sf NodEO} instance can be restarted independently.
\end{itemize}

\begin{figure}[!htb]
\centering
\includegraphics[width=4in]{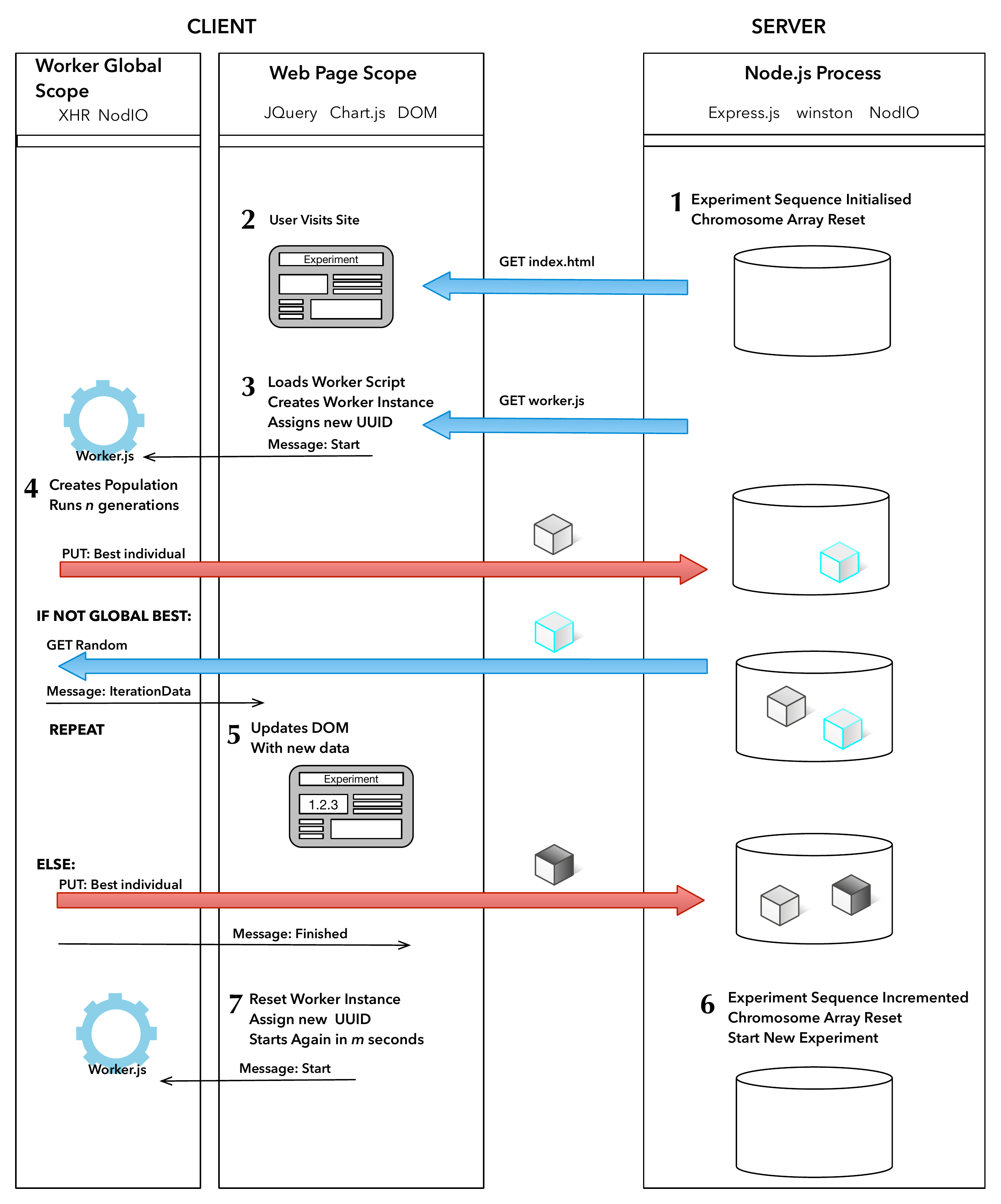}
\caption{Description of the W$^2$ version of {\sf NodIO}. In this
  version, clients use Web Workers to run the evolutionary algorithm,
  with two of them per browser.}
\label{fig:system:w2}
\end{figure}

A sequence diagram of NodIO-W$^2$, is presented
in Figure~\ref{fig:system:w2}. In this version, clients use the Web Workers
API to run two {\sf NodEO} instances per browser. The sequence diagram shows the
interaction between processes in a time sequence. In this version there are
two kinds of asynchronous messages: HTTP Requests shown as bold arrows between
the client and the server where PUT and GET HTTP methods are represented by
red and blue arrows respectively. There are also messages between workers and
the main thread locally in the client, which are shown as thin solid arrows.
There are three processes involved:
\begin{itemize}
\item {\em Node.js} This is the server side process which is basically
  the same as used previously, except for some minor log-related
  changes. 
\item {\em Client main thread} The main script creates
worker instances and handles the user interface.
\item {\em Worker global scope} The worker execution space does not have
access to the Document Object Model (DOM) or JQuery objects in the main thread.
Asynchronous HTTP requests to the server are implemented using the
XMLHttpRequest object.  
\end{itemize}

The sequence of events happens as follows:
\begin{enumerate}
\item First the {\sf NodIO} process is started in Node.js. As part of the
initialization procedure, the sequential number of experiments is initialized and a log
file is created. The shared pool implemented as an array is reset.
\item A volunteer follows the link of the experiment triggering a GET HTTP
request, then the page is loaded.
\item The main script renders the page and creates the worker instance. In
order to create a dedicated worker instance, another GET request is sent to the
server, then the client receives the second worker implementation script.
Each worker instance executes over time one or more evolutionary algorithms
{\em islands}. Each island starts with a random population and it is assigned a
universally unique identifier (UUID). The UUID is included in each of the HTTP
requests sent to the server.
The initial parameters for each island
can be distinct. Once the worker is created and the parameters have been set,
a message is sent to the worker to start the algorithm.
\item The worker first creates the population following the previously
specified parameters. Once the population has been initialized, the EA
generation loop is started. After {\em n} generations the state of the population
is evaluated. The chromosome with the best fitness is sent to the shared
population with a PUT HTTP request. If the best fitness is not the global best,
the EA loop must continue, but first a chromosome is randomly selected from
the shared population. The current generation number and the best
chromosome are also sent to the client's main thread in order to update the DOM to
display the current state of the island (described next). These messages are
all asynchronous, so the EA loop of
the worker continues even before a message is received by the target. If the best fitness is a global best, the best
chromosome is sent to the server and the loop ends. A message is sent to the
main thread indicating that the best individual has been found.
\item The main thread receives messages from workers as events handled by
a callback function. When an iteration message is received, the page is
updated by displaying a dynamic plot of the generation number and best
fitness, and a representation of the chromosome is also generated.
\item When a global best is received from an island, the current experiment
ends, the experiment number is incremented, and the population array is reset.
\item When the main thread receives a message indicating that a worker has ended the EA
loop, that worker is reinitialized. The worker process is not ended, to
avoid the start-up cost, only the parameters and population are reset and a new UUID
assigned.
\end{enumerate}

These two versions of the framework will have to be tested {\em in the
  wild} to measure its performance. 
However, before testing this framework in a real experiment, we have
to establish a baseline for comparison. 

%---------------------------------------------------------------
\section{Baseline Experiments} 
\label{sec:experiments}

We have used the classical Trap function \cite{Ackley1987}
used as well as the Rastrigin's floating-point optimization
problem. Since JavaScript is a functional language and declaring a different 
function and handing at the creation of the algorithm object, called
{\tt Classic}, is the only change needed to work with a different
problem. 

\begin{figure}[!t]
\centering
\includegraphics[width=10cm]{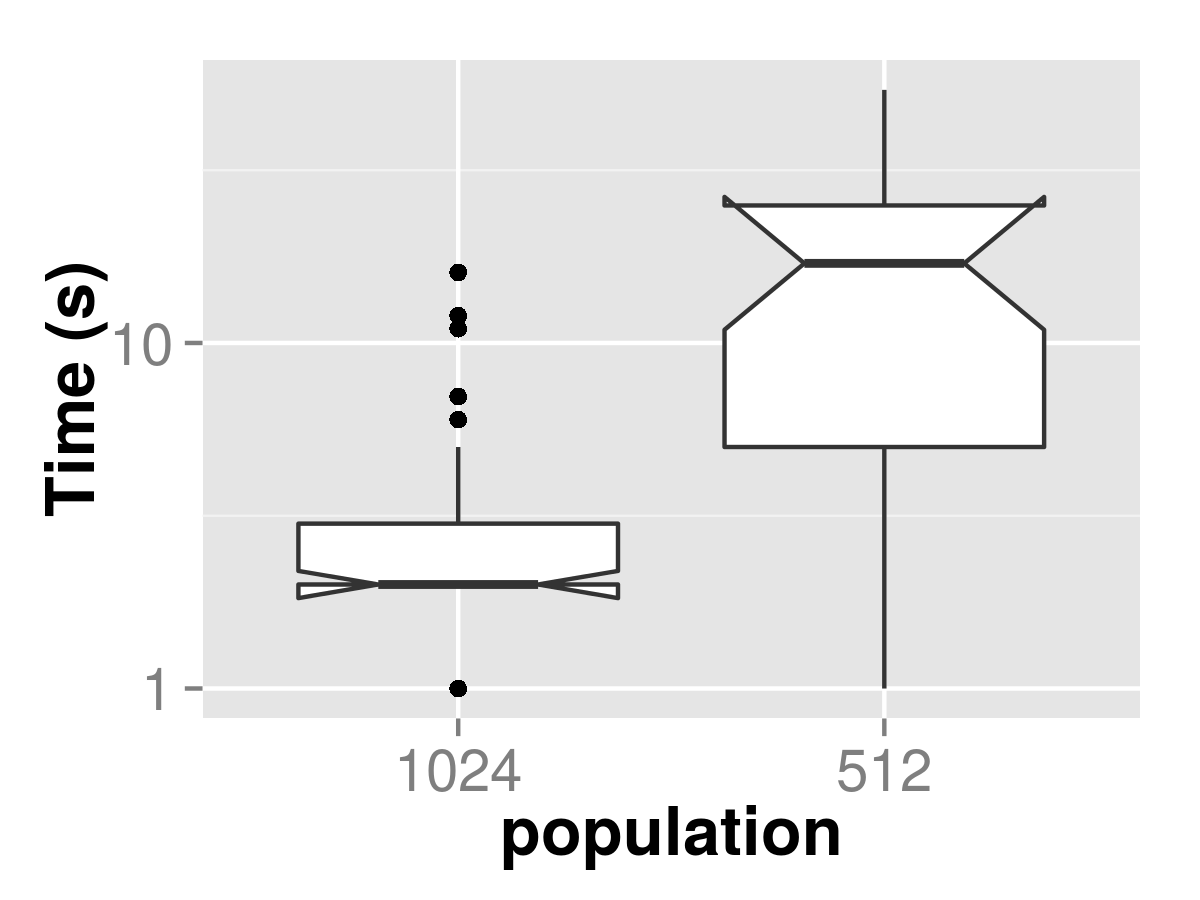}
\caption{Comparison of times to solution (labeled as {\bf times}) for the baseline system, using a {\tt
    Node.js} client, for different population sizes. Only
runs in which solution has been found are used for this graph.}
% Plotted with: ~/Code/splash-volunteer/data/plot-baseline.R
\label{fig:baseline}
\end{figure}
First, we establish the baseline performance by running fifty runs of the
evolutionary algorithm in a desktop client written using {\sf NodEO}
\cite{nodeo2014}, the basic JavaScript library we have used as base to
build {\sf NodIO}, the framework. This experiment tries to 
find the solution to the 40-trap function with parameters $l=4, a=1,
b=2, z=3$ and a population size equal to 512. The algorithm in a
single island was run until the solution (a string with all
ones) was found or five million evaluations had been performed. It
took around a minute, on average, that is $68.9694$ seconds, to
perform these runs. In this experiment only 33, that is, 66\% were
successful. The experiments were repeated for population $p=1024$,
with a success rate upgraded to 100\% and an {\em average} duration of $3.46$
seconds. Results for these two experiments
are shown in Figure \ref{fig:baseline}. The baseline hardware system has these
characteristics: 
\begin{verbatim}
Linux penny 3.13.0-34-generic #60-Ubuntu SMP Wed Aug 13 15:45:27 UTC
2014 x86_64 x86_64 x86_64 GNU/Linux
\end{verbatim}
with a 4-core {\tt Intel(R) Core(TM) i7-4770 CPU @ 3.40GHz}.

These two experiments establish a baseline result and show that
the time to success depends on the population size, with a bigger population
contributing to have more diversity and thus speeding up the solution \cite{DBLP:conf/lion/LaredoDFGB13}. The volunteer computing
experiments that we will describe next do not and can not have the
same conditions, but
the baseline is that if they eventually take longer than a basic
desktop, their interest will be purely academic.

\subsection{Tests with a floating-point hard optimization problem} 

JavaScript has been traditionally implemented as an interpreted
language not designed for the development of high performance
systems. On the other hand, current JavaScript  
virtual machines (VMs from now on) are closing the gap, for instance the Google V8 
is an engine specifically designed for the fast execution of 
large JavaScript applications. In order to increase performance
V8 compiles JavaScript source code directly into machine code when it is first executed. 
There are no intermediate byte codes or an interpreter \cite{Gray:2009:GCM:1610564.1610565}.
JavaScript has gained popularity in server-side development using runtime
environments such as Node.js and is also used in desktop and mobile applications. 
%Mario: Creo que el párrafo de arriba sale sobrando
% (Paloma) Agree, I mean, it adds information but in case you have space problems... is ok if you start the section without it
To further evaluate how a current JavaScript implementation could be used by a
computer scientist to develop complex optimization problems,
a benchmark function, which was provided by the CEC2010 Special Session on
Large-Scale Global Optimization \cite{tang2007benchmark} is used in this section. 
The function is described next.

The basic Rastrigin's function is separable, and is defined as follows;
\begin{equation}
F_{rastrigin}(x)=\sum\limits_{i=1}^D [ x_{i}^{2}-10\cos(2\pi x_i)+10  ] 
\end{equation}
where $D$ is the dimension and $x = (x1, x2, · · · , x_{D})$ is a
$D$-dimensional row vector (i.e., a $1 × D$ matrix). Rastrigin’s function 
is a classical multimodal problem. Such problem is difficult since the number of local
optima grows exponentially with the increase of dimensionality. To make it 
non-separable, an orthogonal matrix is also used for coordinate rotation.
The rotated Rastrigin's function is defined as follows:
\begin{equation}
F_{rot\_rastrigin}(x)=F_{rastrigin}(\textbf{z}), \textbf{z}= \textbf{x} * \textbf{M}
\end{equation}
Where $\textbf{M}$ is  a $D \times{D}$ orthogonal matrix. The benchmark
emulates real-world optimization problems that most likely will consist 
of different groups of parameters with strong dependencies within
but little interaction between the groups. This issue is reflected in the benchmark 
by randomly dividing the objective variables
into several groups, each of which contains a number of variables. The parameter $m$ 
is used to control the number of variables in each group and hence, defining the degree
of separability:
\begin{equation}
F_{15}(x)=\sum\limits_{k=1}^\frac{D}{m}F_{rot\_rastrigin}[\textbf{z}(P_{(k-1)*m+1}:P_{k*m})] 
\end{equation}
For the benchmark $D = 1000$, Group size $m = 50$, $\textbf{x} = (x_1, x_2, \ldots , x_{D})$ is the
candidate solution, $\textbf{o} = (o_1, o_2, \ldots  , o_{D})$ is the shifted global optimum,
$\textbf{z} = \textbf{x} - \textbf{o}$ is the shifted candidate solution and
finally $\textbf{P}$ is a random permutation of $\lbrace1, 2, \ldots  ,D\rbrace$.

This optimization problem was selected because it is representative of the kind of
algorithms, data types, and structures employed in large scale optimization problems. 
The function was implemented in both Matlab and Java languages as part of the
Test Suite of the companion competition of the Special Session and to give competitors an idea of the
computational cost of the challenge, the runtime required for 10,000 function evaluations 
for a particular configuration was published by organizers. The Java
implementation took 7596ms and the Matlab version 1115ms. The whole experiment had the value of
3,000,000 function evaluations as termination condition. Using the Java implementation as
a guide, the function described above was implemented to compare the performance of JavaScript against
both implementations, again using the time required for 10,000 function evaluations. 
Before discussing the setup and performance results, certain details of
the implementation in JavaScript are presented to give readers an idea of 
how practical and mature this language is for developing complex mathematical functions.
\begin{figure}[!htb]
\centering
\includegraphics[width=0.9\linewidth]{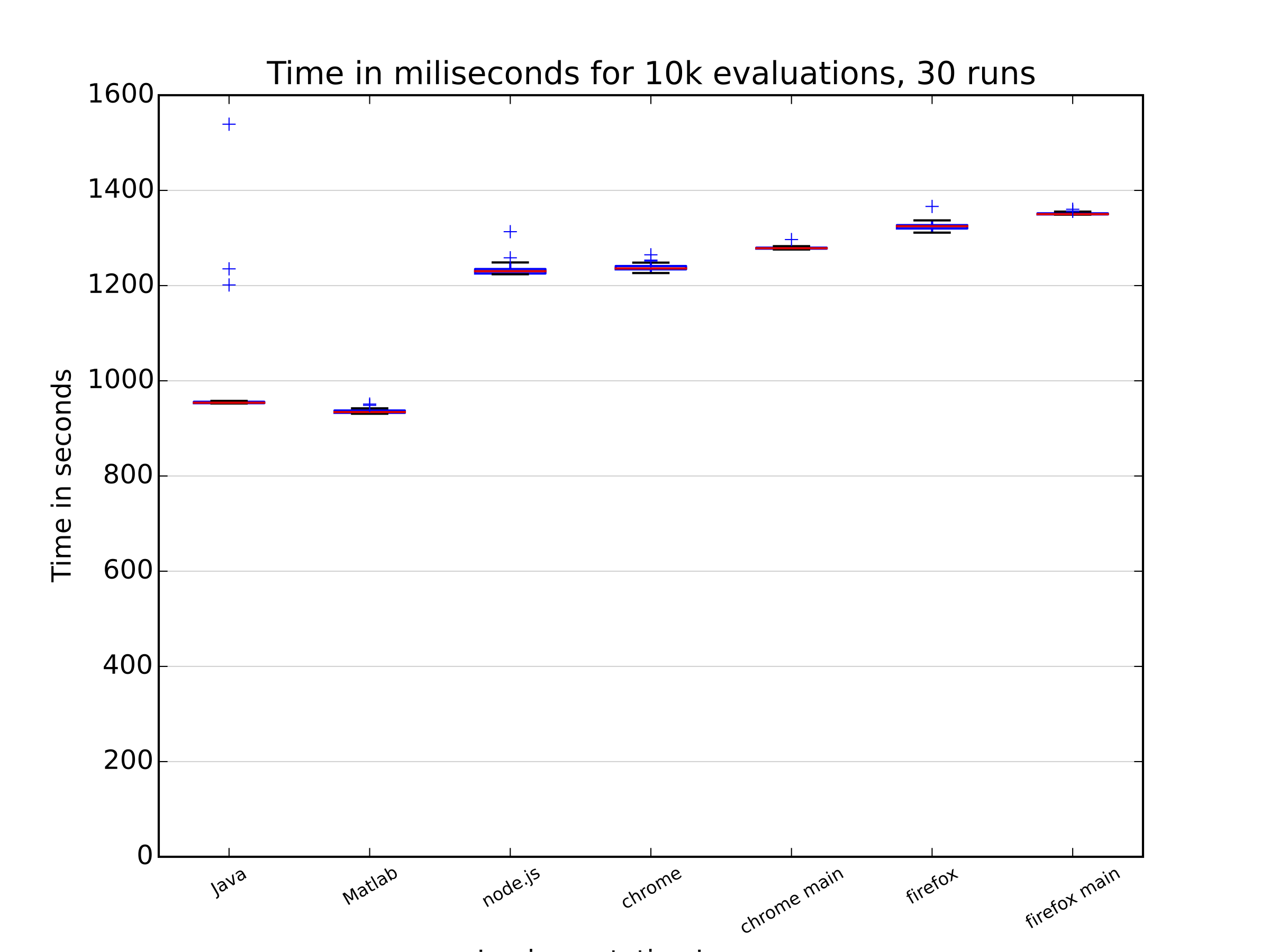}
\caption{ Runtime of 10,000 function evaluations for $F_{15}$} 
\label{fig:f15_times}
\end{figure}

\begin{itemize}
\item {\em Randomize} Both Matlab and Java programs rely on a 
Java Randomizer library for the 
randomization of the shift vectors and matrices used by the function.
The library includes the generation of pseudorandom Gaussians of type
double and uniformly distributed integers. Generators use a single
{\tt long} seed. In order to implement the same functionality in
JavaScript, the library {\tt random-js} was used, as there are inconsistencies 
in the implementation of the standard {\tt Math.random()} between engines
and also their results are non-deterministic. 
The library {\tt random-js}, implements a Mersenne Twister algorithm producing 
consistent results across all JavaScript VMs.
\item {\em Timing functions} In order to compare the performance of 
each implementation, an accurate and consistent way of measuring 
the runtime of functions is needed.
In JavaScript the {\tt Date class} is often used 
to measure execution time, but has the limitation of having a maximum 
resolution given in milliseconds.
In order to measure the runtime intervals with a higher precision two functions
where used: for Node.js we used the native {\tt process.hrtime()} function
which returns a high-resolution measure in a {\tt [seconds, nanoseconds]}
array, and for measuring in the browser the {\tt Performance.now()} function was called,
which uses floating-point numbers with up to microsecond precision. 
A drawback found is that the {\tt Performance.now()} function is only implemented
in Firefox and Chrome browsers. Both functions are independent of the system clock. 
\item {\em Tools} An important consideration when choosing certain 
language to write non-trivial programs is the availability of development tools. While developing
the experiments, we used a comprehensive set of developer tools available for both
browsers and desktop \cite{tilkov2010node}. These tools included
package managers, in this case {\tt npm}, debuggers, logging libraries
such as {\tt winston}, 
network monitors, editors, and even debugging of the multi-threaded execution of Web Workers.
Although the fact that developer tools vary from browser to browser
supposed a drawback, they all have the same functionality. The
JavaScript debugger can be accessed directly from the user menu in
most browsers. 
\item {\em Data Types} JavaScript uses floating point numbers with a 
limited precision of 64 bits that partially implement the functionality of
the {\tt StrictMath} library used by the Java implementation. If more precision
is needed, developers could use {\tt math.js}, an extensive math library with
support for matrices and big numbers.
\end{itemize}

We found that the JavaScript ecosystem is mature enough for developing non-trivial 
code, but developers have to consider the differences between VMs implementations
and therefore use certain libraries in order to increase the precision and repeatability of
algorithms. As for the performance of JavaScript against Java and Matlab
we conducted first the basic test described earlier in a 3.7 GHz Quad-Core
Intel Xeon E5 processor running OS X 10.10.5, 
using Java(TM) SE Runtime Environment (build 1.8.0\_25-b17), 
MatLab version R2015a for Mac OSX, Node FrameWork version 0.12.2,
Google Chrome Version 46.0.2490.86 (64-bit) 
and Firefox version 41.0.2. % (Paloma) Table here? Or at least the characteristics could be represented the same way as for the baseline experiment
 The performance of the time required for 10,000 function evaluations is presented in
Figure \ref{fig:f15_times}. In our test the Matlab implementation had the best
performance with an average of 935ms, followed by Java with
991ms. On the other hand, JavaScript took 32\% more time than Java with 1238ms in Chrome using a 
single Web Worker and 1,234ms using Node.js. % Are these results correct? JS=>1240ms and Node.js=>1,270ms ???
The results also show that there is not much difference between the Node.js
implementation and that of a
browser. There is also not much difference between running the code in the main thread or in Web
Workers. When running the experiment in Firefox a dialog alerted that the
script was taking too much time, showing that this
kind of processing is not intended for the main thread. Additionally, there is not much overhead
when running two
Web Workers in parallel as they took 1279ms each. This experiment
shows that JavaScript is viable as a language
language for developing complex mathematical functions an these can run in a browser
with acceptable performance.

%---------------------------------------------------------------
\section{Conclusion}
\label{sec:conclusion}

This paper has been intended mainly as a description of the {\sf
  NodIO} architecture as well as two different implementations, with
or without web workers, as well as a collection of measurements of the
performance of the JavaScript language in different implementations,
and compared with other languages.

The first conclusion we can draw is that, even if JavaScript is not
the fastest language around, its performance is competitive and can be
used profitably for medium size problems, with its speed in floating
point problems being around 30\% less than Matlab or Java. 

A framework created around this language, like the {\sf NodIO} we
systematize and introduce in this paper, can more than compensate this
difference with the recourse of the users that can be spontaneously
gathered in a volunteer computing experiment. The fact that it is so
simple to participate in it, without needing to download anything,
implies that it can be used for peak or non-permanent workloads.

However, it remains to be seen how many users and the real performance
that {\sf NodIO}, in any of its implementations, would achieve. This
is left for future work.

%---------------------------------------------------------------
\section*{Acknowledgment}

This work has been supported in part by TIN2011-28627-C04-02 and
TIN2014-56494-C4-3-P (Spanish Ministry of Economy and Competitivity),
SPIP2014-01437 (Direcci{\'o}n General de Tr{\'a}fico) and PYR-2014-17
GENIL project (CEI-BIOTIC Granada). Additional support was recieved by
Projects 5622.15-P (ITM) and  PROINNOVA 2015: 220590 (CONACYT).
We would also like to thank the
anonymous reviewers of previous versions of this paper who have really
helped us to improve 
this paper (and our work) with their suggestions. We are also grateful
to Anna S\'aez de Tejada for her help with the data processing scripts.

\bibliographystyle{IEEEtran}
\bibliography{geneura,volunteer,javascript,ror-js,GA-general}

\end{document}